\documentclass[12pt,preprint]{aastex}

\shorttitle{Dust Distribution in Gas Disks II: self induced ring formation 
through a clumping instability}
\shortauthors{Klahr \& Lin}

\begin{document}


\title{Dust Distribution in Gas Disks II:\\
    Self Induced Ring Formation Through a Clumping Instability}


\author{Hubert Klahr\altaffilmark{1} and D.N.C.\ Lin}
\affil{UCO/Lick Observatory, University of California,
    Santa Cruz, CA 95064}

\email{VERSION: Wednesday, June. 22nd 2005}
\email{STATUS : ApJ in press}


\altaffiltext{1}{present address: Max-Planck-Institut fuer Astronomie, Heidelberg, Germany}


\begin{abstract}
Debris rings of dust are found around young luminous stars such as
HR4796A and HD141569.  Some of these entities have sharp edges and
gaps which have been interpreted as evidence for the presence of
shepherding and embedded planets.  Here we show that gaps and sharp
edges in the debris disks of dust can also be spontaneously self
generated if they are embedded in optically thin regions of gaseous
disks.  This clumping instability arises in regions where an
enhancement in the dust density leads to local gas temperature and
pressure increases.  Consequently, the relative motion between the gas
and the dust is modified.  The subsequent hydrodynamic drag on the
dust particles leads to further enhancement of their concentration.
We show that this process is linearly unstable and leads to the
formation of ring-like structures within the estimated life time of
such young objects. Once the gas is removed (e.g.\ by photo evaporation)
the structures are ``frozen'' and will persist, even when the
gas might not be observable anymore.
\end{abstract}




\section{Introduction}

The coplanar orbits of nearly all planets in the Solar System inspired
Laplace to propose that they are formed in a gaseous disk, the
primordial solar nebula, centered around the protosun.  Protoplanetary
disks are found around a large fraction of classical and weak-line T
Tauri stars (Stauffer, Prosser, Hartmann, \& McCaughrean 1994).  The
existence of these disks is generally inferred from the excess in the
infrared continuum which is primarily due to dust reprocessing of the
stellar radiation.  These signatures fade with the age of the stars.
In young stellar clusters, the fraction of stars with clear signs of
infrared excess declines with their age on the time scale of a few Myr
(Haisch, Lada, \& Lada 2001).  Around some post T Tauri stars such as
$\beta$ Pic, debris disks have been found. The amount of dust which
emits infrared radiation in these debris disks also appears to be
inversely proportional to the age of their host stars (Zuckerman, Kim,
\& Liu 1995).  This apparent evolutionary pattern may be interpreted
as coagulation of dusty material into relatively large particles
(McCabe {\it et al.} 2003) which are less efficient in reprocessing
the stellar radiation (Yorke \& Sonnhalter 2002).  It is also possible that the
residual grains are depleted from the circumstellar disks by photo
evaporation, stellar wind, or some dynamical processes.  In either
case, the decrease in the infrared excess casts a constraint on the
magnitude of time scale for planet formation to be in the range of a
few million years.

Current searches for extra solar planets indicate that they are
present in at least eight percent of all target stars (Marcy, Butler,
\& Vogt 2000).  As data accumulate with increasing time bases, the
discoveries of long period planets are anticipated around a much
larger fraction of all planets (Trilling, Lunine, \& Benz 2002,
Armitage {\it et al.} 2002, Ida \& Lin 2003).  If so, some debris
disks may indeed contain newly formed protoplanets since they
represent advanced evolutionary stages of protostellar disks.  These
planets are expected to tidally interact with their ambient gas and
dust by exciting density waves (Goldreich \& Tremaine 1982, Papaloizou
\& Lin 1984), inducing gaps with sharp edges (Borderies, Goldreich \&
Tremaine 1989, Lin \& Papaloizou 1986, Artymowicz, \& Lubow 1994,
Bryden, Ro'zyczka, Lin, \& Bodenheimer 2000), confining rings between
them (Bryden, Chen, Lin, Nelson, \& Papaloizou 1999), and warping the
disk (Larwood, \& Papaloizou 1997).

The discovery of a debris disk around HR4796A was particularly
interesting because its sharp edges resemble that expected for tidally
truncated disks (Schneider {\it et al.} 1999).  Recent HST images of
another disk around HD141569 clearly indicate that its outer region is
strongly perturbed by its binary companion (Clampin, Ford,
Illingworth, Krist \& Ardila in preparation).  But, this disk appears
to be composed of two concentric rings which are separated by a narrow
gap (Weinberger {\it et al.}  1999).  The binary star's tidal torques
occur on length scales comparable to the disk radius and are unlikely
to induce the formation of the fine structure seen in this disk.  A
more plausible and potential exciting alternative culprit is a
Jupiter-mass planet which may be embedded within the gap (Weinberger
{\it et al.}  1999).  But, the distance of the hypothetical planets
from their host stars in both cases are more than an order of
magnitude larger than that of the gaseous giant planets in the solar
system.  The dynamical time scale in these disk regions exceeds $10^3$
yr.  Also in these regions, the disk may attain sufficient mass to
become gravitationally unstable and remain optically thin.  The
formation of giant planets through gravitational instability may be
rapid (Boss 1998) if the fragment can contract and collapse before
being sheared out (Pickett, Mejia, Durisen, Cassen, Berry, \& Link in
press; Pickett, Durisen, Cassen, \& Mejia 2000; Nelson, Benz, Adams,
\& Arnett 1998). The necessary condition for a rapid contraction is
that the cooling time scale must be comparable to or shorter than the
local dynamical time scale (Johnson \& Gammie 2003).

Application of the conventional core-accretion theory of planet
formation suggests that the formation time scale for Jovian-mass
planets at these large distances from the central object is much 
longer than the
age of HR4796A and HD141569 even under the most favorable conditions
(Pollack {\it et al} 1996).  But the discovery of short-period planets
(Mayor \& Queloz 1995) supports the scenario that planets may migrate
over extensive distance shortly after their formation (Goldreich \&
Tremaine 1980, Ward 1986, Lin \& Papaloizou 1986, Lin, Bodenheimer, \&
Richardson 1996).  A particularly effective mechanism for rapid and
extensive migration is through a runaway interaction between embedded
planets and their nascent (Masset \& Papaloizou 2003, Artymowicz in
preparation) although the termination conditions remain uncertain.
Another potential avenue for extensive outward migration for
protoplanets formed within a few AU from their host stars is through
distant resonant encounter between several massive planets (Lin \& Ida
1997).  In the context of the Solar system, Thommes, Duncan, \&
Levison (2002) suggested that Uranus and Neptune may have migrated
outwards as a consequence of the gravitational interaction between the
outer planets and residual planetesimals. In general, dynamical
instability induces the planets to attain large eccentricities.  The
gap width induced on the nearby disk by a highly eccentric planetary
perturber may be much larger (Artymowicz \& Lubow 1996) than that
observed between the rings around HD 141569.  But, the eccentricity of
the perturber may also be rapidly damped by the collective response of
the residual planetesimal disk (Agnor \& Ward 2002).

The above discussions indicate that despite its potentially profound
implication on the theory of planet formation, the planet-disk
interaction scenario is highly uncertain.  Thus, it is useful to
explore some alternative explanations for the observed features in the
debris disks.  In a previous paper \citep{klahr01}, we suggested that
the debris disks and rings may be confined with relatively sharp edges
through their interaction with both the stellar radiation and residual
gas \citep[sell also][]{takeuchi01}.  Due to a negative pressure
gradient in the radial direction, the azimuthal velocity of gaseous
disks is generally sub Keplerian everywhere.  In opaque regions of the
disk, the grains' motion is affected only by the central star's
gravity and their interaction with the ambient gas.  As they attempt
to attain Keplerian orbits, the grains experience a head wind and the
hydrodynamic drag resulting from this differential motion leads them
to undergo orbital decay (Weidenschilling 1977).

But, in the optically-thin regions of the disk where the grains are
directly exposed to the stellar radiation, the azimuthal velocity of
the grains may also be reduced by the outwardly pointing radiation
pressure from the central star.  Around relatively luminous stars,
such as HR4796A and HD 141569, the grains' azimuthal velocity may be
less than that of the gas such that the combined effects of the
radiation pressure and hydrodynamic drag can induce the grain to
migrate outward.  The above effects require the presence of residual
gas which is otherwise difficult to detect directly.  For example, at
large distances from its central star, CO gas may be depleted as
volatile molecules are condensed onto the grains (Thi {\it et al.}
2002).  The detection of molecular hydrogen around some of these
debris disks have been reported (Thi {\it et al.} 2001), although
observational uncertainties remain (Sheret, Ramsey-Howat, \& Dent 2002).

In the context of the debris disk around HR 4796A, we showed that due
to these hydrodynamic drag and radiation pressure, a slight
inhomogeneity in the gaseous background can lead to the formation of
the dusty rings with sharp edges.  Thus, the observed features in the
ring do not necessarily imply the presence of embedded planets.
Furthermore, if these sharp features are indeed an indicator for some
residual gas, a less stringent formation time scale would be required
for the formation of gaseous planets in the outskirts of planetary
systems (for the formation of Neptune, see Bryden {\it et al.}  1999).

Nevertheless, gas may have been present at the time 
when the structures in HR4796A and HD 141569 formed.
After the ring structures were created by assistance
of the gas, the gas may well have been evaporated and
depleted to a level below observability and below a
value that would still influence the dust. The dust 
rings would then be the ``frozen'' final state of
the ring formation process.

In this paper, we extend our previous investigation on the dynamical
evolution of optically thin debris disks around relatively luminous
stars.  In our previous work, we assumed an arbitrary surface density
distribution for the gas.  Our main objective here is to illustrate
that the feedback interaction between the gas and dust naturally leads
to a clumping instability in which the dust drifts into an
equilibrium configuration until the dusty rings become self confined.  We show
that this clumping instability results naturally and the structure of
the debris disks can be determined self consistently.  In \S2, we
outline our basic assumptions and formulate the evolution of both dust
and gas. A linear stability analysis in \S3 is followed by some
numerical models in \S4 and their observational signature in \S5. We
summarize our results and discuss their implications in \S6. In the
appendix we give a simple estimation for realistic $\beta$ values
(i.e.\ the sensitivity of gas temperature on grain density).

\section{Basic assumptions for the model}
We consider an optically thin debris disk where the gas dust plasma
is being irradiated by stellar light.  The gas attains an equilibrium
temperature somewhere between a maximal and a minimal possible
temperature.  The maximum temperature is given by the blackbody
temperature of the dust grains and the minimum temperature by the
ground state of the gas molecules and the temperature of the
interstellar space.

\subsection{Gas Temperature}
The equilibrium temperature of the gas $T_g$ in an optically thin regime
at low temperatures is given by the balance between radiative cooling
via line emission and heating due to various effects including
photo-electric heating via the silicate grains, the gas molecules' inelastic collisions with the dust grains as well as line absorption of UV
photons (Kamp \& van Zadelhoff 2001; Ferland, Korista, Verner, Ferguson, Kingdon \& Verner 1998; Woitke, Kr\"uger \& Sedelmayr 1996). 
The maximum temperature which can be attained by the
gas is already given by the temperature of the dust grains, which
themselves are heated to blackbody temperature by irradiation from the
central star. In the current analysis, we are primarily interested in
the interaction between the gas and the dust.  We neglect effects such
as super-heating of small grains due to radiative inefficiencies
(Chiang \& Goldreich 1997) for this analysis as it is not important to
understand the basic instability.

In general, the gas temperature is a complex function of radius $R$,
luminosity of the central object $L_{*}$, surface of the dust grains,
surface density of the gas $\Sigma_g$, surface density of the dust
content $\Sigma_d$, the cross section of the dust $A_d$, and most
of all the complete chemical composition of the gas $N_n$ and the
ionization state of the molecules $X_n$, determining the possible line
transitions leading to the emission of radiation:
\begin{equation}
T_g = f(R, L_*, A_d, \Sigma_g, \Sigma_d, N_n, X_n, ...).
\end{equation}
The detailed determination of the function $f$ will be presented
elsewhere.

For the present purpose of examining the stability of dust
to gas interaction, the most important parameter is 
\begin{equation}
\beta \equiv {\partial {\rm ln T} \over \partial {\ln \Sigma_d} }.
\end{equation}
In order to illustrate the evolution of small disturbances, we adopt a
simple generic analytic prescription in which we assume that
\begin{equation}
T_g = T_0 \left(\frac{\Sigma_d}{\Sigma_0}\right)^\beta,
\label{eq:tg}
\end{equation}
where $\beta, T_0$ and $\Sigma_0$ are to be determined later by
radiation transfer models.  In the linear stability analysis, only the
instantaneous slope $\beta$ needs to be specified and the above
explicit prescription for $T_g(\Sigma_d)$ is not needed.
 
The power index $\beta$ gives the steepness of the gas temperature
dependence on the dust load.  In regions where the dust to gas surface
density ratio, $\Sigma_d/\Sigma_g > Z_\odot$, where $Z_\odot$ is the
solar metallicity, the disk gas is primarily heated by conduction to
the gaseous molecules from the dust particles, which themselves are
heated by the irradiation of their host stars and via photo-electric heating 
via the silicate grains.  Qualitatively, $T_g$
is expected to be an increasing function of $\Sigma_d$ because in the
optically thin region of the disk, the fraction of stellar radiation
absorbed by the dust is proportional to the optical depth of the dust.
The rate of conduction between the gas molecules and the dust grains
as well as the number of photo-electric electrons
is also an increasing function of $\Sigma_d$.  Thus, we consider
primarily the range of parameter space for which $\beta >0$ (see
appendix).  However, in the low $\Sigma_d/\Sigma_g$ limit, the gas
molecules are mostly heated directly by the UV photons of their host
stars such that the energy deposition rate to the disk is insensitive
to the magnitude of $\Sigma_d$ (Kamp {\it et al.} 2003). Also in the
opaque regions of the disk, $T_g = T_d$ and  $T_g$ can be independent of
$\Sigma_d$ so that $\beta =0$.  Therefore, we also consider other
values of $\beta$.  Under unlikely circumstances where $\beta < 0$,
the debris disks are stable against this clumping process.

\subsection{Radial Drift velocity}
In our previous analysis (Klahr \& Lin 2001), we derived the radial
drift velocities $v_r$ for particles under the combined influence of
non-Keplerian gas motion and radiation pressure. (The velocity of the
dust is expressed in lower case $v$'s whereas that of the gas is
expressed in capital letters $V$'s.) For small particles with radius
$a_d$ and density $\rho_d$ that are in the Epstein regime the
friction time is $\tau_f = a_d \rho_d/(\rho_g c_s)$ in a gas of
density $\rho_g$ at a sound speed of $c_s$. The radial velocity is then
\begin{equation}
v_r = \tau_f 2 \Omega^* dV^* \,\,\,\,\,\,{\rm for}\,\,\,\,\,\, St << 1,
\label{Eq_v_r}
\end{equation}
where the Stokes Number $St = \tau_f \Omega^* << 1$, $\Omega^*$ is the
particles' hypothetical orbital frequency subjected only to the stars'
gravity and radiation pressure. The difference of the actual orbital
velocity of the disk gas and the hypothetical particle velocity
$\Omega^* R$, $dV^*$, can have either positive or negative signs
depending on the structure of the gas disk.

Similarly, the radial drift velocity of the larger particles can be
expressed as
\begin{equation}
v_r = dV^* \,\,\,\,\,\,{\rm for}\,\,\,\,\,\, St = 1,
\label{Eq_v_r1}
\end{equation}
and
\begin{equation}
v_r =  \frac{2  dV^*}{\Omega^*\tau_f}\,\,\,\,\,\,{\rm for}\,\,\,\,\,\, St >> 1.
\label{Eq_v_r2}
\end{equation}
We utilize this expression for the radial drift velocities in the
linear stability analysis below.

\section{Stability Analysis}

The evolution of an axis-symmetric  dust distribution is determined by the 
continuity equation in cylindrical coordinates:
\begin{equation}
{\partial \over \partial t} \Sigma_d + 
\frac{1}{r}{\partial \over \partial r} r \Sigma_d v_r = 0.
\end{equation}
In general one would have to solve a continuity equation for all
single dust species as their radial drift velocity depends on the
friction time. Such detailed approach will be carried out in future
models which will try to fit the observations in detail.  For
this work we assume a mono dispersive dust distribution which is
characterized by one particle size, shape and friction time.

The radial drift velocity for the most mobile particles ($St = 1$) is:
\begin{equation}
v_r = dV^*,
\label{Eq_v_r3}
\end{equation}
and $dV^* = V_\phi - v^*_\phi$ is a function of the particles theoretical
azimuthal velocity $ v^*_\phi$  for circular orbit in the absence of gas
\begin{equation}
v^*_\phi = \sqrt{\frac{G M_\star-\frac{L_\star A_d}{4 \pi m_d c}}{r}}.
\label{eq:nearKep}
\end{equation}      
Here $A_d$ is the cross section and $m_d$ the mass of the dust grains.
The speed of light is denoted by $c$.

We want to stress that the characteristic radial drift velocity will
also be correct within an order of magnitude for particles which are
up to 20 times larger or 20 times smaller than the characteristic size.

The azimuthal velocity of the gas is $V_\phi$ and is determined by the
radial pressure gradient in first oder approximation to
\begin{equation}
V_\phi = \Omega r +  \frac{1}{2 \Omega \Sigma_g}{\partial \over
\partial r} {P}
\end{equation} 
where $\Omega = (G M_\ast / r^3) ^{1/2}$ is the Keplerian frequency.
The pressure $P$ is given by an ideal gas equation of state a function
of temperature and gas surface density $\Sigma_g$ thus:
\begin{equation}
V_\phi = \Omega r +  \frac{R_{gas}}{2 \mu \Omega}\left(\frac{T_g}{\Sigma_g} 
{\partial \over \partial r} {\Sigma_g} + {\partial \over \partial r}
{T_g}\right)
\label{eq:vphi}
\end{equation}     
where $R_{gas}$ is the gas constant.  Note that the second and third
terms on the right hand side of eq.\ (\ref{eq:vphi}) are due to the
surface density and temperature distribution of the gas respectively.
In general, $\Sigma_g$ is not perturbed by the dust
concentration.  But any clumping in the dust would lead to a local
enhancement in $\Sigma_d$ which in term heats the gas.  Because the
temperature distribution is much more sensitive than the $\Sigma_g$
distribution, we focus our stability analysis on those contributions
which are proportional to the temperature gradient.

Combining all the equations into one expression and defining 
a local mean sound speed to be $\frac{R_{gas} T_0}{\mu} = c_s^2$,
we find:
\begin{equation}
{\partial \over \partial t} \Sigma_d = - \frac{1}{r}{\partial \over \partial r}
r \Sigma_d  \left[\Omega r +  \left( {c_s^2 \over 2 \Omega r} \right) 
\left( {T_g \over T_0} \right)  \left( {\partial {\rm ln} {\Sigma_g} \over 
\partial {\rm ln} r} + \beta {\partial {\rm ln} {\Sigma_d} \over 
\partial {\rm ln} r} \right) - \sqrt{\frac{G M_\star-\frac{L_\star A_d}
{4 \pi m_d c}}{r}} \right].
\label{num_eq0}\end{equation}  
With a prescription for $T_g$ such as that in eq(\ref{eq:tg}), this continuity
equation can be solved numerically.

For the linear stability analysis, we simplify the equation to clarify
the mechanism by dropping all terms not important for the instability.
We neglect the effects of radiation pressure which gives only a
systematic offset but no instability. We also assume that the gas is
much more smoothly distributed than the dust, {\it i.e.} $\vert
\partial {\rm ln} \Sigma_g / \partial {\rm ln} r \vert << \vert
\partial {\rm ln} \Sigma_d / \partial r \vert $ so that:
\begin{equation}
{\partial \over \partial t} \Sigma_d = - \frac{1}{r} {\partial \over \partial r}
\left( \left(\frac{T_g}{T_0}\right)\frac{c_s^2 \beta } {2 \Omega} {\partial 
{\rm ln} {\Sigma_d} \over \partial {\rm ln} r} \right).
\label{num_eq} 
\end{equation}

Linearized $\Sigma_d = \Sigma_0 + \Sigma_d^\prime$, we find in a local
approximation with $\frac{1}{r}\partial_r r \rightarrow \partial x$
at a given radius $r = R$
\begin{equation}
{\partial \over \partial t} \Sigma_d^\prime + \beta\frac{c_s^2}{2 \Omega} 
{\partial^2 \over \partial x^2} {\Sigma_d^\prime} = 0.
\end{equation}
The local approximation does not limit the physical validity
of our analysis. It simplifies the mathematics but does not change the
underlying physics. Using a polar coordinate
system would have required to apply e.g.\ Bessel functions instead of Fourier modes. 
The result would have been the same but it would be less transparent.
However, $(1/r \partial_r r = \partial_x)$ requires $x/r <
1$. In the case of HR4796A the radial extent of the ring is 20 AU
where the radius is 70 AU, so the approximation is fulfilled in this system.

For short-wavelength disturbances, the above equation has a solution
$\Sigma_d ^\prime \sim e^{-i\left(\omega t - k x\right)}$.  The
growth-rate of these disturbance is:
\begin{equation}
\Gamma = -i \omega =  \beta\frac{c_s^2}{2 \Omega}  k^2.
\end{equation}
If we replace $c_s = H \Omega$ and define $n = k R$ then the above equation
reduces to 
\begin{equation}
\Gamma = -i \omega =  \frac{1}{2} \beta\left(n \frac{H}{R} \right)^2 \Omega.
\label{eq:growth}
\end{equation}
This result indicates that a $\Sigma_g$ variation can lead to local
heating of the gas which in term modifies the local pressure gradient
and the azimuthal speed of the gas.  Through a hydrodynamic drag, the
dust particles respond by clumping together to enhance the local
$\Sigma_g$.  This clumping instability is related by, but not the same as
the diffusion and thermal instability in gaseous accretion disks
(Pringle 1981). In both cases, the growth of a perturbation is due to
the efficiency of angular momentum transfer being inversely
proportional to the surface density.  But the present clumping
instability involves a different physical process which requires the
feedback between the gas and the dust through the radiative heating of
the dust and the local heating of the gas via the dust.

It is interesting to notice that the growth rates are close to the
orbital frequency for $\frac{H}{R} \approx \frac{1}{n}$.  In this
limit, the instability will grow on a dynamical time scale. 
For disturbance with $n > R/H$ the
linear leading-terms approximation is inappropriate.  
On these scale, the thin disk approximation breaks
down and the gas pressure in the direction normal to the plane has a
tendency to stabilize the flow.  In addition, very small scale
variations, if unstable can reverse the gradient of angular momentum in
the disk (see eq.\ \ref{eq:vphi}).  A negative angular momentum
gradient, even localized can lead to interchange instabilities which
results in mixing in the radial direction.  Since we are primarily
interested in identifying the condition for the onset of a clumping
instability, we shall present elsewhere the growth limit due to
amplitude saturation and nonlinear effects.

Nevertheless, eventually the gas will be removed from
the system (e.g. photo evaporation). As a consequence
the friction time of the particles 
will increase and the radial drift 
velocity decrease (see eq.\ \ref{Eq_v_r2}).
Thus, the right hand side of eq(\ref{num_eq}) will 
eventually vanish. This means that the structure is
now in first order approximation constant in time.
So even the gas is removed from the system the
structure in the dust distribution will remain.
Consequently we can not argue that there must still be 
gas at a dynamically important or even at an observable 
amount in HR4796A and HD141569.

Note that $\Gamma$ and $\beta$ have the same sign.  It is not
forbidden that $\beta$ may attain a negative value, in which case the
perturbation would be damped.  However, this case requires the gas
temperature to decrease with $\Sigma_d$ which seems to be physically
unlikely for an optically thin debris disk. Only when the dust grains
are the efficient cooling agent and the heating of the gas is provided
by other means then $\beta$ might become negative. Yet this case 
is unlikely for the objects that we consider.

In opaque regions of a circumstellar disk where all
radiation is intercepted by the dust, the gas temperature becomes
independent of $\Sigma_d$ so that $\beta=0$. In these $\beta=0$ cases,
the perturbation remains neutral and the disk is stable.

\section{Numerical Solution}
In order to have an illustration of how the instability can evolve
into the non linear region, we obtain numerical solutions of the
complete continuity equation \ref{num_eq0} without substituting
the gas temperature $T_g$ and assuming circular orbits:
\begin{equation}
{\partial \over \partial t} \Sigma_d = - \frac{1}{r} {\partial
\over \partial r}
\left( r^{-5/2} \frac{\Sigma_d}{\Sigma_g} {\partial
\over \partial r} \Sigma_g T_g \right).
\end{equation}
Here we set all physical constants to one and used $\Omega =
r^{-3/2}$. Using these dimensionless units sets the unit time scale to
one orbital period at the radius $r$ times the inverse square of the pressure scale height. 
\begin{equation}
\tau_0 = \left(\frac{H}{r}\right)^{-2} \frac{2 \pi}{\Omega}
\end{equation}
Thus, it will be simple
to calculate the physical time scale for a real object like HR4796A 
afterwards by simply multiplying the dimensionless time by the unit time scale of the physical
radius, that is considered.

 The gas temperature is calculated from the power-law
prescription for $T_g$ as expressed in eq.\ (\ref{eq:tg}), where we
additionally limited the dynamical range to one order of magnitude:
\begin{equation}
T_g = T_0 \max{\left[3.33,\min{\left[0.33,
\left(\frac{\Sigma_d}{\Sigma_0}\right)^\beta\right]}\right]}.
\label{eq:tg2}
\end{equation}
As we will discuss in the appendix the gas temperature
will probably stay in such limits (see Figure \ref{fig_6}). 

The analytic estimate in eq.\ (\ref{eq:growth}) indicates that the growth
rate is proportional to $n^2$, {\it i.e.} the most unstable
perturbations are those with the smallest wave numbers.  We have
already indicated above that the thin disk approximation for the
continuity equation breaks down for $n > R/H$.  In reality, the
vertical pressure gradient becomes important to stabilize the
disturbance.  We also argued that the non linear growth of disturbance
with $n > R/H$ would lead to the violation of the Rayleigh's criterion
for axis-symmetric rotational flows.  If such a situation arises, mixing
would almost certainly limit the growth of the clumpy structure.  In
order to avoid the unphysical growth on scales smaller than $H$, we
introduce a prescription by adding a stabilizing contribution 
to the
continuity equation such that
\begin{equation}
{\partial \over \partial t} \Sigma_d = - \frac{1}{r} {\partial
\over \partial r}
\left( r^{5/2} \frac{\Sigma_d}{\Sigma_g} {\partial
\over \partial r} \Sigma_g T_g  - H^2 {\partial \over \partial r} {\partial^2 \over 
\partial r^2} \Sigma_d\right).
\end{equation}
With this prescription, the equation can now be discretized with
central differences\footnote{The necessity for the stabilizing contribution arises from
the fact that we use a central differencing scheme. In future
projects we will adapt an upwind advection scheme which is stable by construction even without the artificial stabilization.} on a grid with 100 radial points and 2 ghost cells
on each side, where $\Sigma_d$ was set to fixed values $\Sigma_{inner}
= 1.0$ and $\Sigma_{outer} = 10^{-10}$. 

 Another choice of boundary conditions would have been
vanishing gradients. But this leads to numerical problems
with the scheme we applied, i.e.\ negative densities can occur close
to the boundary. The application of fixed values for the ghost cells
helps to avoid this, whereas the overall ring structures
are not changed in size and location.

Once one starts to model a given observation, one has to include
more physics than we do now. Then one needs the detailed cooling
and heating behavior for the optical thin gas. One will then
also find that only a certain region around the star will be subject
to the clumping instability. For instance at very large radii the
gas will be at a lowest possible temperature of maybe 10K which
will be determined by the back ground radiation field, thus there
is no feed back mechanism from the dust density to the 
temperature of the gas.
At radii to close to the star dust might be destroyed by
high temperatures and also no such instability is possible.
In reality the radial window for instability might be even
narrower than that, because the cooling behavior depends on
the chemical composition of the disk gas, which itself
depends on the radiation field. Only where the right 
cooling agents are present one can hope that the cooling
behavior gives rise to the clumping instability. 
Detailed investigation on real objects are necessary to
identify the potentially ring forming regions.

   In such a system, where the instability is truncated 
towards the edges, one will have more natural boundary
conditions. In our simplified model we thus assume 
that the dust density does not change at the edges,
which has a similar effect as shutting of the instability
there.

We use a
numerical scheme using a fourth order Runge Kutta integrator to
advance the distribution in time. 
The fixed background gas density distribution was chosen
to be $\Sigma_g \propto r^{-3/2}$. A similar profile was chosen for initial distribution of the dust $\Sigma_d$ with a randomly fluctuation of $1\%$
amplitude.  We also set $T_0 \propto r^{-1/2}$ with $H= 0.1$.

The stabilizing contribution is the only limit for the growth 
of the shortest modes in this flat approximation where vertical structure is neglected.
A more realistic modeling would require to resolve the 3D structure
of the gas disk and the turbulent processes within. Nevertheless, 
this is way beyond the considerations of this paper. On top of this we would
only shift the application of free parameters one layer up, 
without gaining much additional insight. 

A pressure change can lead to the modification in the density distribution in
both vertical and radial direction.  Since hydrostatic equilibrium is
maintained in the vertical direction by a balance between pressure and
gravity, a change in the pressure can modify the density distribution
above the mid plane.  But it does not necessarily change the surface
density.  In the radial direction, equilibrium is maintained by a balance
between centrifugal force of the Keplerian rotation and the stellar gravity.
Thus, small variations in the pressure do not modify the gas density
significantly.

The linearized solution in eq(\ref{eq:growth}) clearly indicates that
the sign of $\beta$ determines the growth and damping of the
disturbances.  We choose three values of $\beta$ (-1, 1, 2) to
illustrate this dependence.  The magnitude of $\beta$ also modifies
the radial location of the dust concentration maxima. 

\begin{figure}
\plotone{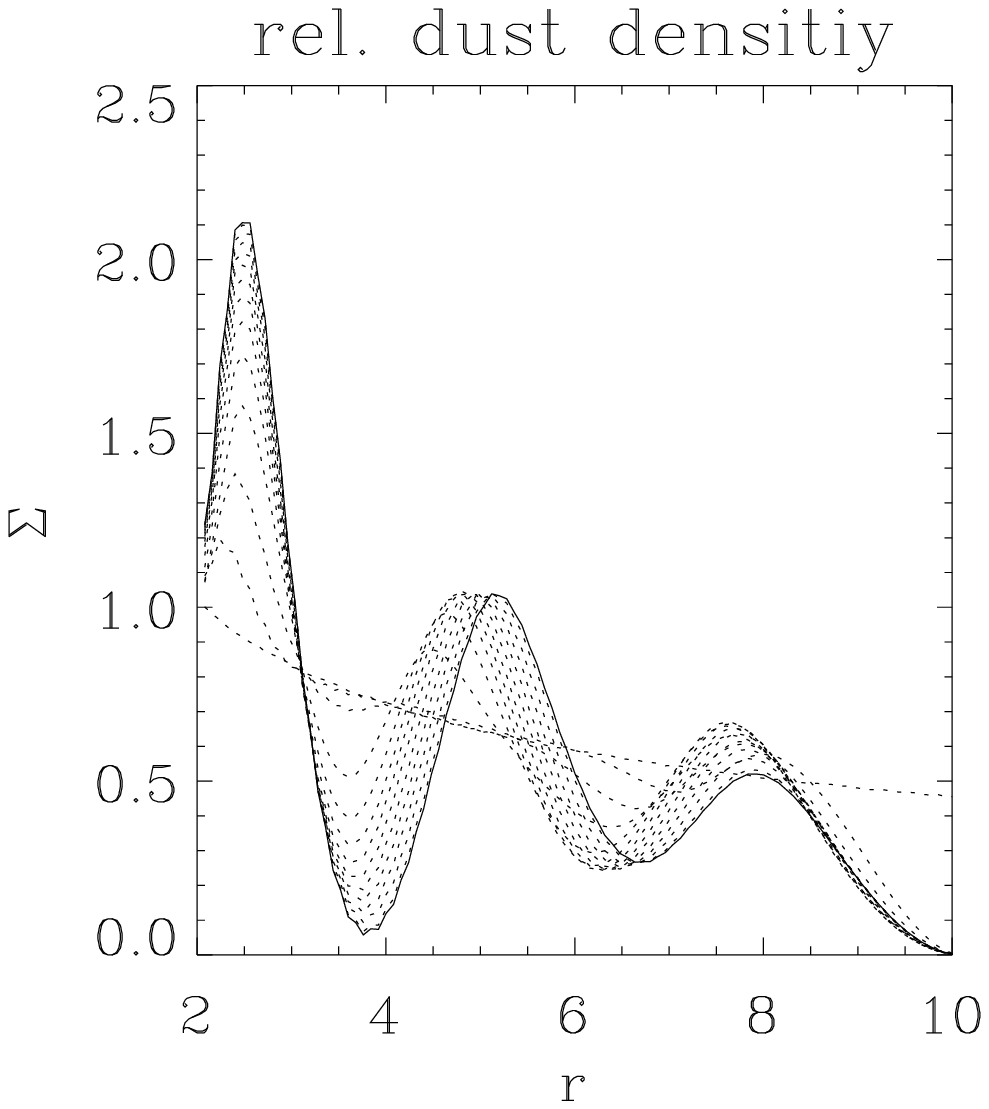}
\caption{Evolution of an initially homogeneous dust distribution over dimensionless time $t=1$ in units of the initial value over the dimensionless radius. $\beta = 1.$\label{fig_1}} 
\end{figure}

Fig.\ \ref{fig_1} shows the evolution of a small random $\Sigma_d$ 
fluctuation with $\beta = 1$. 
The solid line denotes the final time, where we stopped the
integration at the dimensionless time $t_{max}= 1$.
The density is only fixed in the ghost zones, which we do include in the plot,
thus the values close to the boundary may differ from the fixed values (see above).  
The doted lines are snapshots after evenly
spaced periods of time. We plotted the
surface density of the dust
$\Sigma_d(t)$ over the dimensionless radius.

   The instability is such an effective process that
the initial state does not matter. With more effort we could have
started with a smoother distribution towards the edges, but the result 
would have been the same.

The results in fig.\ \ref{fig_1} clearly
indicate the growth of the perturbed $\Sigma_d$ and the formation of
ringlets which are separated by gaps.  The width of the rings is
an increasing function of the radius because $H/r$ increases with $r$.
Eventually, as the amplitudes of $\Sigma_d$ are strong enough, 
the residual disk attains an equilibrium in which the
residual dust rings become isolated.

The numerical models needed 1 dimensionless time scale to develop rings.
The mean radius of the simulation was chosen to be $r=5$ thus 
it took only $0.1$ dimensionless orbits at the center of the dust density ring.
This typical time scale can be derived from eqs.\ (12) and (19)
to be: $\tau_0 = \left(\frac{H}{r}\right)^{-2} \frac{2 \pi}{\Omega}$. 
If we apply parameters estimated for HR4796A
(see Model A in Klahr and Lin 2001) we can estimate the
time the instability needed for growth at the ring location of 
$70$AU. The orbital period is $370$ yrs. The pressure scale 
height is determined from the irradiation to be $H/R = 0.07$.
It follows that the typical time scale is only $7.5 \times 10^4$ yrs,
which means the instability developed in the model after about
$10^4$ yrs for the most efficient drifting particles. 
This model (Model A from Klahr and Lin 2003) assumes
the existence of gas at an amount of $100 M_\earth$ in the disk around
HR4796A. With such a high gas density dust grains of $600 \mu$ in size
are drifting radially at the highest possible rate $(\Omega \tau_f \approx 1$ see eq.\ \ref{Eq_v_r3}).
The radial drift velocity decreases with the friction time and thus with gas density.
In Klahr and Lin (2001) we also assumed two models (Models C and D) that contain only
10 $M_\earth$ respectively 1 $M_\earth$ in the gas. Under such conditions the instability needs
10 respectively 100 times longer. For HR4796A this means that the ring may have
formed within $10^4$ and $10^6$ yrs depending on the amount of gas that 
was available. Thus, if we assume that the gas was depleted on a time scale
of $10^6$ yrs it is possible to generate the ring structure within this 
time and also to conserve or fixate the structure by removing the gas
during the process.

\begin{figure}
\plotone{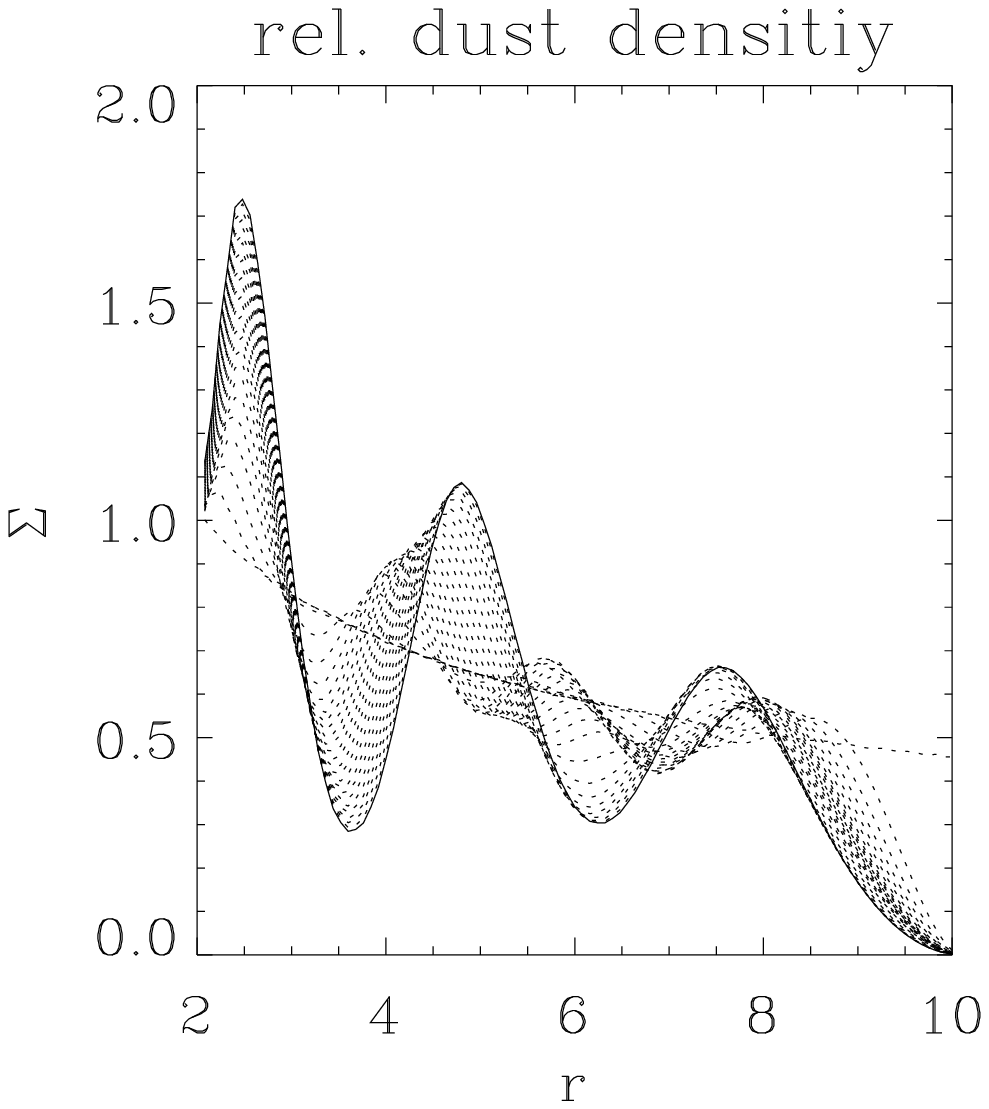}
\caption{Evolution of an initially homogeneous dust distribution over dimensionless time $t=1$ in units of the initial value over the dimensionless radius. $\beta = 2.$\label{fig_2}} 
\end{figure}

\begin{figure}
\plotone{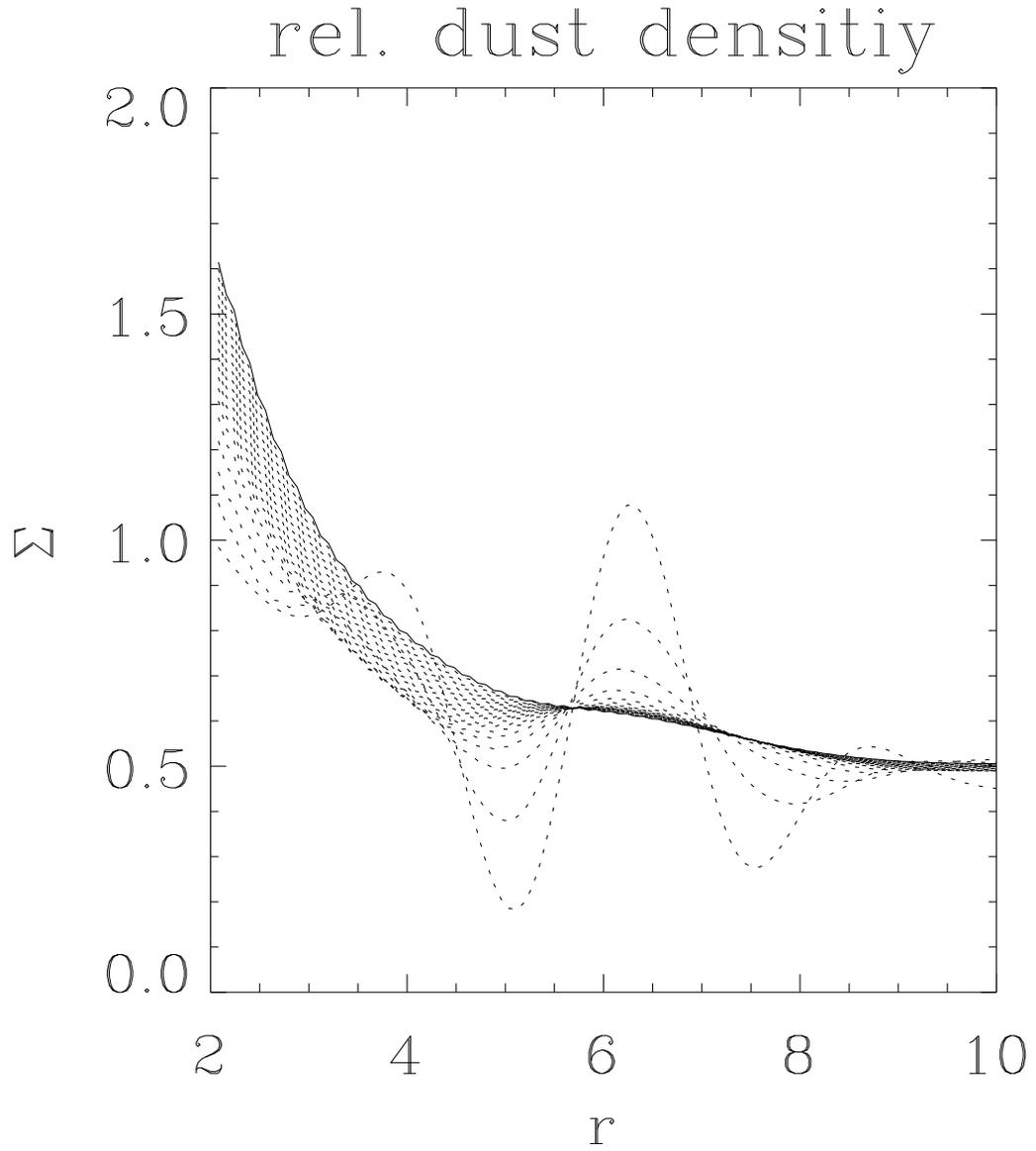}
\caption{Evolution of an initially sinusoidal dust distribution over dimensionless time $t=10$ in units of the initial value over the dimensionless radius. $\beta = -1$.\label{fig_3}} 
\end{figure}

In Figure \ref{fig_2} we present the evolution of $\Sigma_d$ for the $\beta = 2$
model. Maximum time was also $t_{max}=1$. 
The growth of the perturbation in this case is similar to the
$\beta=1$ model though the growth rate is faster and the outward drift
of the rings is more pronounced.  Finally, we present in Figure \ref{fig_3}, the
evolution of $\Sigma_d$ for the $\beta = -1$ model. As this process
is generally slower we integrated until $t_{max}=10$. 
As we have
anticipated in our analytic results that an initially strong sinusoidal 
perturbation is damped in this case.  

\section{Observational Appearance}
\begin{figure}
\plotone{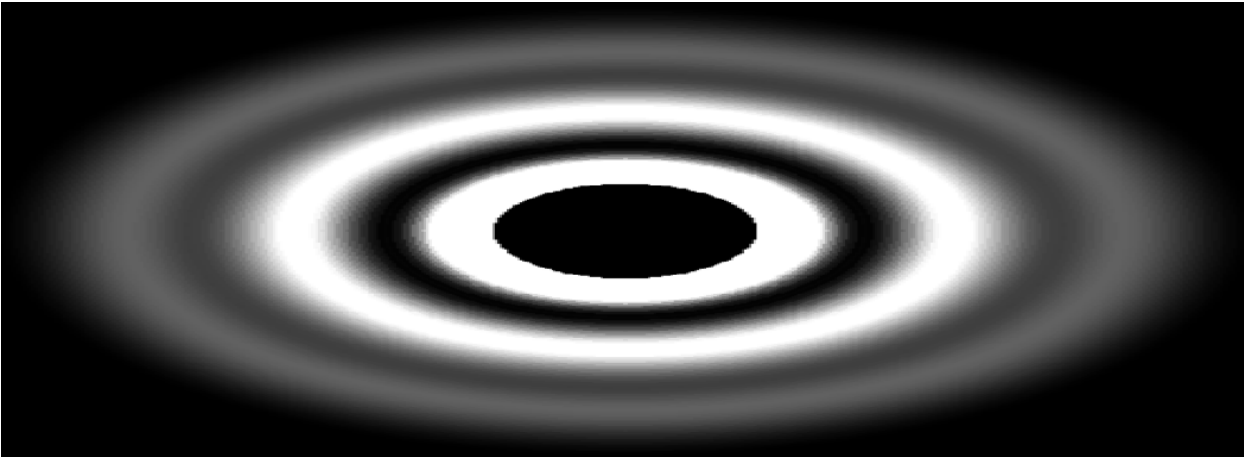}
\caption{Simulated intensity map for an optically thin debris disk. $\beta = 1$.\label{fig_4}} 
\end{figure}

\begin{figure}
\plotone{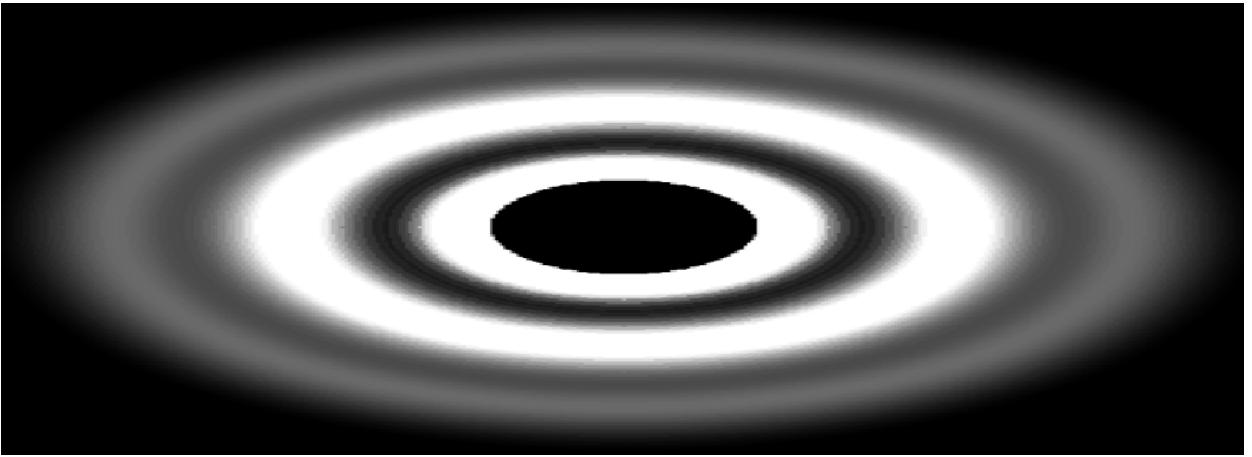}
\caption{Simulated intensity map for an optically thin debris disk. $\beta = 2$.
Now the middle ring is wider than in the previous case.\label{fig_5}}
\end{figure}

The radial dust density and temperature distribution 
can be translated in an intensity map via
\begin{equation}
I(r,\phi) \propto \Sigma_d T_d^4.
\end{equation}
In figures \ref{fig_4} and \ref{fig_5} we show these maps for the $\beta=1$ and  $\beta=2$
case. In the latter case the middle ring is a little wider
than in the previous case. One should not expect to much difference
between both cases, as $\beta$ predominantly influences the time scale
on which the rings form but not the structures themselves.

The resemblance to the observations of HR4796A and HD141569 is
striking. Future work could focus on a possible instability 
of non-axisymmetric modes, which are actually already observed.

\section{Conclusion}

In this paper, we consider the interaction between dust, gas, and
radiation in debris disks around young stellar objects.  In general,
the gas and the dust component do not have the same azimuthal speed.
The gas attains a dynamical equilibrium in which the centrifugal force
is balanced by the central stars' gravity and a negative pressure
gradient in the disk.  The azimuthal speed of the gas is generally sub
Keplerian.  In a gas-free optically-thin environment, the dust attains
a dynamical equilibrium in which the centrifugal force is balanced by
the central stars' gravity and a radiation pressure.  In a gaseous 
environment, hydrodynamical drag also influences the motion of the gas
and dust.  In most regions of the disk where the dust concentration
is relatively small, the gas causes the dust to migrate inward/outward
if the azimuthal speed of the gas is smaller/larger than that of the 
dust.  

However, in relatively massive Herbig Ae/Be stars, the UV intensity is
relatively strong.  Gas may be rapidly depleted through photo
evaporation process (Shu, Johnstone, \& Hollenbach 1993, Hollenbach
{\it et al.} 1994).  Nevertheless, a small amount of residual gas and
dust may be retained around systems such as HR4796A and HD141569 for
a few Myr.  Regions with modest gas density and dust opacity are 
primarily heated by the absorption of stellar irradiation on the dust
grains.  The gas is indirectly heated through photo-electric heating
via the dust grains.

We consider the evolutionary stability of a two-fluid flow under these
circumstances.  For simplicity, we consider the limiting case that the
azimuthal speed of the gas is smaller than that of the dust initially.
(Similar results can be obtained in the opposite limit).  We find that
in some optically thin regions of the disk, the gas temperature is an
increasing function of the surface density of the dust.  In this case,
a slight concentration of dust leads to local pressure maxima in the
gas.  The positive pressure gradient interior to these maxima
increases the local azimuthal velocity of the gas which remains
smaller than that of the dust initially.  Although the dust continues
to experience a hydrodynamic drag and migrate inward, the strength of
the head-wind is weaker and the inward radial velocity is smaller.
Similarly, exterior to these maxima, the pressure gradient is more
negative than elsewhere so that the azimuthal speed of the gas is
slower than its surrounding region.  This enhanced head wind causes
the dust to migrate toward the maxima at a faster pace.  The
congestion caused by the slow down in the radial migration of the dust
on the inside and the speed up on the outside of the surface density
maxima in turn leads to an increase in $\Sigma_d$.  This process leads
to a clumping instability.

The initial growth of the clumping instability is similar to the
viscous diffusion instability in accretion disk models.  But, in
contrast to the viscous instability, the rings can evolve into
apparently isolated entities. As the amplitude of the surface density
variation increases, the modification of the local pressure
distribution intensifies.  Eventually, the azimuthal speed of the gas
decreases below that of the dust in some region interior to the
surface density maximum.  The dust particles move outward as they
experience a tail wind. This inflow barrier leads to the apparent
isolation of the rings, while the gas distribution is merely unchanged
during the entire process, but the gaseous component of the residual
disk is difficult to be detected anyway.  Although this self shepherding
effect diminishes with gas depletion, the diffusion of the grains due
to the gas drag effect also vanishes.  After the residual gas is
completely evaporated, the ring structure would be preserved until
particle collisions and radiation drag induces its spreading.  But
these processes occurs on a much longer time scale than a few Myr.

The growth rate of this clumping instability is inversely proportional
to the radial wavelength of the perturbation so that it leads to the
formation of concentric rings with gaps between them.  The lower limit
on the width of the ring is determined by the local pressure scale
height.  In accordance with the Rayleigh's criterion, rings with
narrow widths are dynamically unstable.  Although our discussion is
based on the assumption that the azimuthal velocity of the gas is
initially smaller than that of the dust, the same argument applies for
the opposite limit.  In that case it is the congestion of the dust's
outward flow which causes the growth of the clumps.

Debris disks have been seen around many post T Tauri stars.  Some of
these systems such as $\beta$ Pic have relatively smooth surface
density distribution in the radial direction while those around
HR4796A and HD141569 appears to have ringlet structure.  While it is
tempting to interpret the ringlet structure with a tidal
interaction between the residual disk and some embedded planets, our
results indicate the existence of alternative scenarios which can also
account for the sharp features in the ring.

The stability condition is determined by the functional dependence of
$T_g$ on $\Sigma_d$.  The instability arises when the gas temperature
is an increasing function of the dust surface density.  Such
conditions are not satisfied in the optically thick regions of the
disk or in regions where the gas molecules radiate either much more or
much less efficiently than they are heated by conduction.  If gas is
an efficient radiator, $T_g$ would attain the value that corresponds to the
ground state of the gas molecules.  In the opposite limit, gas would
attain the radiative equilibrium temperature of the grains.  In both
limits, the gas temperature becomes independent of $\Sigma_d$,
$\beta=0$, and the debris disk is stable.  It is also possible that
$\beta$ may attain a negative value though it seems physically
unlikely for a debris disk. 
We shall examine elsewhere the functional dependence of the
gas temperature on other physical parameters.  This analysis will
include a more realistic dust size distribution and the radiation
transfer of the gas.  We will also provide detailed modeling of the
observations. It is entirely possible that only over some special range
of disk radii the clumping instability may arise.

\acknowledgments
We want to thank Inga Kamp for her thoroughly checking
of the heating and cooling properties of optically thin disks. 
This work was supported by NASA TPF program under the grant NNG04G191G.
Work supported in part by the European Community's Human Potential
Programme under contract HPRN-CT-2002-00308, PLANETS.

\appendix
\section{Gas Temperature Estimate}
For this work we assumed a temperature versus dust density
relation with a positive slope, without further discussing
the exact shape of this relation. A detailed exact derivation 
would need the use and modification of a radiation transfer
code (see Kamp \& van Zadelhoff 2001; Ferland, Korista, Verner, Ferguson, Kingdon \& Verner 1998; Woitke, Kr\"uger \& Sedelmayr 1996).
This shall not be done here.
 
   But we can derive some general expressions for
the gas temperature ($T$) as it depends on the surface
area that is provided by the dust grains.
The strongest heating $Q^+$ provided to the
Hydrogen molecules probably occurs by photo-electric heating 
via the silicate grains (Kamp \& van Zadelhoff 2001):
\begin{equation}
\Gamma_{pe} = 2.5 \times 10^{-4} \sigma \epsilon \chi n_{\rm H}.
\end{equation}
Here $\sigma = \frac{n}{n_{\rm H}} \sigma_d$ is the total
grain cross section per H nucleus, $\epsilon$ is the photo-electric 
efficiency, and $\chi$ is the intensity of the UV field from the
central object ($\sigma_d$ is the cross section and $n$ the
number of dust grains). 

Actually $\epsilon$ also depends on the gas temperature,
but for the expected low temperatures $T_g << 10^4$K this dependence
is negligible for our purpose. $\epsilon$  depends on $n_e$ and $\chi$ 
and hence using reasonable values of those two, e.g. $n_e = 1 {\rm cm}^{-3}$, $\chi = 300$
at 70 AU, $\epsilon$ can only vary as much as by a factor 2
between 10 and 100 K.

\begin{figure}
\plotone{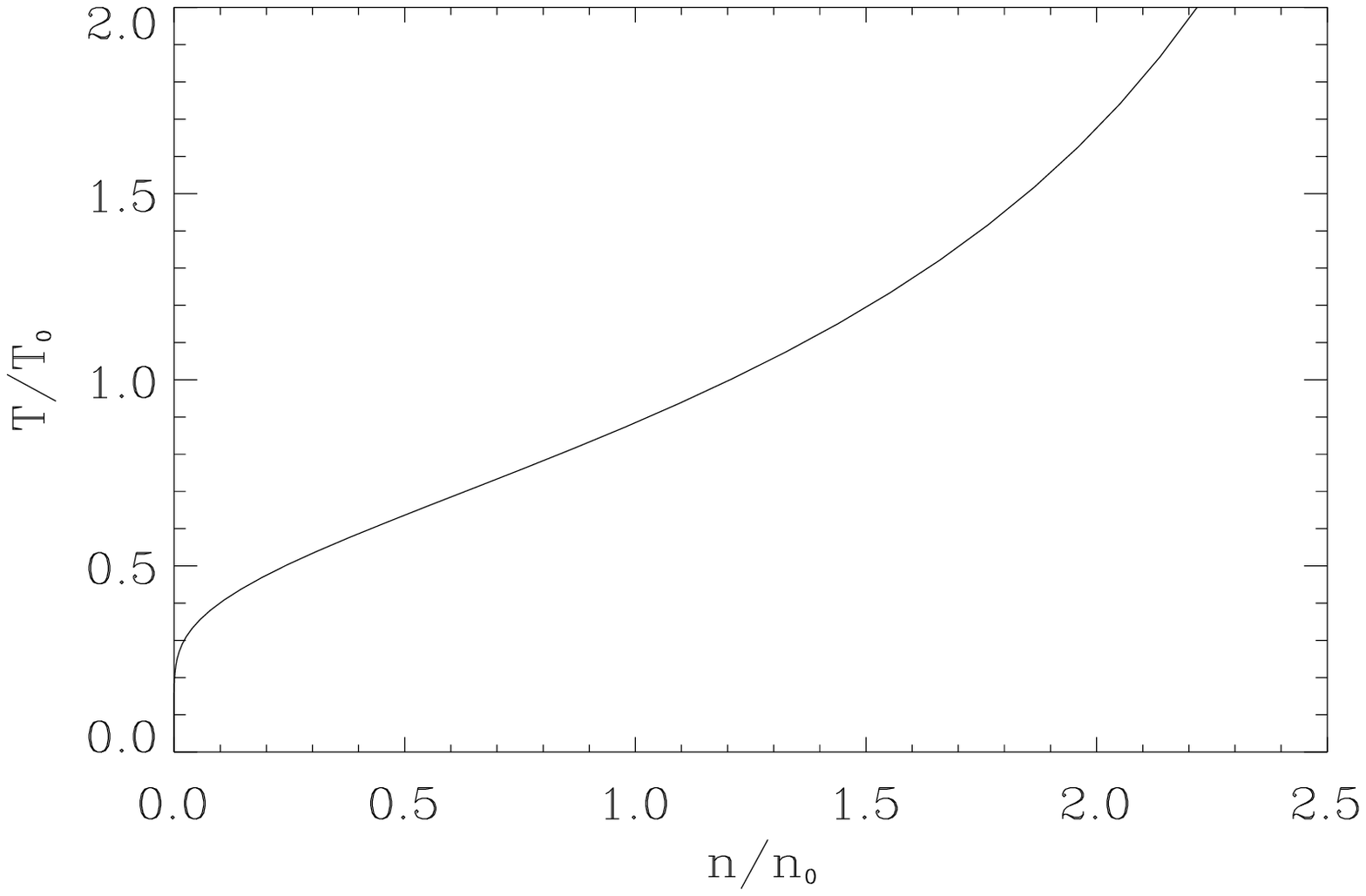}
\caption{Gas temperature in dependence on the number of dust grains. See text for details.\label{fig_6}} 
\end{figure}

Thus it follows, that the heating is in first order
linear increasing with the number of the dust grains:
\begin{equation}
\Gamma_{pe} \propto n.
\end{equation}
The Hydrogen being at its ground state can not cool 
via radiation but only via collisions with other atoms or molecules.
These collision partners themselves will cool via rotational 
 or fine structure transitions (Woitke et al.\  1996).
Given the strong UV radiation field of a young A0\,V star, the
dominant cooling agent is probably ionized carbon ${\rm CII}$ (Kamp {\it et al.} 2003).
One can expect that C is totally ionized and $n_C+ \propto 10^{-4} n_{\rm H}$.
We use the two level approximation of Hollenbach \& McKee (1989)
for the ground state of the ionized carbon: It is the 
157.5 $\mu$ line that has a cooling rate of:
\begin{equation}
\Lambda_2 = f_1 n({\rm C II}) A_{10} h\nu_{10}.
\end{equation} 
The fraction of ${\rm C II}$ in the upper level can be calculated via
\begin{equation}
f_1 = \frac{g_1 e^{-E_1/kT}}{g_0+g_1 e^{-E_1/kT}},
\end{equation} 
with the statistical weights $E_0 = 0.0$ and $E_1 = 1.27 \times 10^{-14}$ and 
$g_0 = 2. g_1 = 4$ respectively. $A_{10}$ is the Einstein probability for
spontaneous emission.
When the heating is balanced by cooling, we can calculate
the dependence of equilibrium temperature on the number of dust grains:
\begin{equation}
\frac{g_1 e^{-E_1/kT}}{g_0+g_1 e^{-E_1/kT}} \propto \frac{n}{n_{\rm H}}
\end{equation}
which can be easily solved for $T$.
In Fig.\ \ref{fig_6} we plot the dependence of $T/T_o$ on $n_/n_{0}$.
The exact values for $T_0$ and $n_0$ have to be calculated
model dependent, but we already see that $\beta$
in this plot is roughly $1.0$ (which is independent on the explicit model) 
and we conclude that the
instability can occur on reasonable time scales.




\clearpage










\end{document}